\documentstyle[aps,pre,multicol,graphicx]{revtex}

\newcommand{\uprule}{\end{multicols}
\noindent \vrule width3.375in height.2pt depth.2pt 
\vrule height.5em depth.2pt \hfill \widetext }
\newcommand{\downrule}{\indent \hfill \vrule depth.5em height0pt 
\vrule width3.375in height.2pt depth.2pt 
\begin{multicols}{2} \narrowtext}

\makeatletter

\renewenvironment{figure}
  {\let\@capwidth\linewidth\def\@captype{figure}}
  {}
\makeatother

\begin{document}
\draft
%

\title{Study of a generalized Metropolis decision rule in auxiliary field
quantum Monte Carlo}
\author{C. L. Martin\cite{martin_email}}
\address{Physics Department, University of California, Santa Barbara, California 93106}
\author{R. M. Fye\cite{fye_email}}
\address{Sandia National Laboratories, MS-1111, Albuquerque, New Mexico 87185}
\date{\today}
\maketitle

\begin{abstract}
  We consider a generalization of the standard Metropolis algorithm
  acceptance/rejection decision rule and numerically explore its
  properties using auxiliary field quantum Monte Carlo.  The
  generalization involves a free parameter which, given
  a criterion for proposing attempted moves, can be used to tune the
  average acceptance rate in a particular way.
  Such tuning can also potentially change Monte Carlo autocorrelation
  times, and the combination of the changing acceptance rate and
  autocorrelation times raises the possibility of more efficient
  simulations.  We explore these issues using primarily massively
  parallel quantum Monte Carlo runs of the ``test case''
  two-dimensional Hubbard model, and discuss results and applications.
\end{abstract}



\begin{multicols}{2}

\section{Introduction}

Since its introduction\cite{1953:Metro.Rosen.Rosen.Telle},
the ``Metropolis algorithm'' for Monte Carlo
has become a widely used and powerful numerical tool, both for classical
and quantum systems.
The algorithm describes a Markov chain through states of the system
with rules for first proposing a new state and then
deciding whether or not to accept a move to the proposed state.
In statistical mechanics simulations, the Metropolis
rules by construction sample states according to their Boltzmann weights,
allowing computation of expectation values.
However, this Boltzmann weight sampling
can be satisfied by other ways of making the
acceptance/rejection decision besides the conventional one.
We consider here a class of such alternative
acceptance/rejection decisions
and explore whether they
might be used to increase simulation efficiency.
We focus in this paper on auxiliary-field fermion quantum Monte Carlo
simulations\cite{1981:Scala.Sugar,1981:Blank.Scala.Sugar,1981:Scala.Sugar*1,1983:Hirsc,1985:Hirsc,1986:Hirsc.Fye,1988:Fye.Hirsc,1989:White.Scala.Sugar.Loh,1992:Loh.Guber}
though our approach and results may have more general
application, particularly in other electron quantum Monte Carlo
methods \cite{1994:Hammo.Leste.Reyno,1995:Ander,1996:Ceper.Mitas,1996:Mitas,1998:Night.Umrig,1999:Umrig2}.
We use primarily massively parallel machines in the quantum Monte Carlo
simulations, which has allowed us to more easily gather
data than would otherwise be possible.

One way to increase general Monte Carlo efficiency is to minimize the
autocorrelation times $\tau_{e}$.
$\tau_{e}$ is qualitatively defined for
an observable as the number of
Monte Carlo steps over which measurements of that observable remain
correlated once the system has equilibrated (i.e.,
once the system has become independent of the initial configuration).
$N$ Monte Carlo steps then correspond to approximately
$N_{\rm ind} = N / \tau_{e}$
independent measurements. A particular statistical error will require a given
number $N_{\rm ind}$ of uncorrelated data points. Hence, $N = N_{\rm ind}
\, \tau_{e}$
steps will be required for a given desired statistical error,
with reducing $\tau_{e}$ commensurately reducing Monte Carlo
run time. A related consideration involves the ``warm-up'' or
equilibration time $\tau_{w}$,
a measure of the average number of steps required to go from typically
low-weight initial
configurations to the high-weight region of the phase space where
measurements can then be taken.

In auxiliary field and other determinantal
quantum Monte Carlo methods, however, another factor is introduced with regards
to efficiency: an accepted move is often
so much more computationally intensive than a
rejected move that the time required for rejected moves can be
neglected compared to the computationally dominating
acceptance calculations
\cite{1981:Scala.Sugar,1981:Blank.Scala.Sugar,1981:Scala.Sugar*1,1983:Hirsc,1985:Hirsc,1986:Hirsc.Fye,1988:Fye.Hirsc,1977:Ceper.Chest.Kalos}.
Then, the dominant
computational time required for a simulation of $N$ steps
is proportional rather to
$N A_{e} = N_{\rm ind} A_{e} \tau_{e}$, where $A_e$ is the
average equilibrated move acceptance rate,
so that one now wishes to minimize the quantity $A_{e} \tau_{e}$.
Analogously, the dominant required computational
time for equilibration will be
$A_{w} \tau_{w}$, where $A_{w}$ is an
acceptance rate averaged during the warm-up process.
Exploring the above efficiency issues is a major focus of the paper.
One aspect of this involves the calculation of equilibration and
autocorrelation times for the ``test case'' two-dimensional Hubbard model,
which results, to the best of our knowledge, have not previously appeared.

After the introduction, we discuss various Monte Carlo
algorithmic considerations,
including parallelization issues, a
generalization of the conventional Metropolis acceptance/rejection rule,
and relevance to auxiliary field and other determinantal quantum Monte Carlo.
We then explore properties of the new Metropolis generalization using primarily
massively parallel
auxiliary field quantum Monte Carlo
simulations of the two-dimensional Hubbard model. 
Lastly, we discuss the numerical results and summarize.

\section{Monte Carlo Algorithm Considerations}

\subsection{Parallelization Issues}

Currently, the fastest computers available are those with a
massively parallel architecture.
These computers consist of hundreds to thousands of individual processors
linked so that they can communicate with each other.
Efficient use of these machines is governed by
keeping communication time sufficiently low that computational
speed scales roughly linearly with the number of processors.

One approach for using such computers in Monte Carlo simulations is to
distribute a single Monte Carlo ``walk''
over all the processors.  This allows the
computational work of each step in the walk to be spread out over all the
processors, but it can require a high degree of communication.
This approach can be feasible for classical simulations
of large systems with short-range forces, with each processor ``owning''
part of the system. However, it has not been shown to be efficient for 
determinantal quantum Monte Carlo simulations
of the type we will discuss, and in particular might be expected to be
less efficient for two dimensional systems.

A second approach is to let every processor perform its own independent
Monte Carlo ``walk''.
Each processor then starts with a different initial configuration and
can proceed
independently of the other processors, and no
communication is required until the final accumulation of results
\cite{1994:Bonca.Guber,1994:Fye,1995:Scale.Runge.Lee.Corre.Oklob.Vujic}.

A potential drawback to this second approach, however, is that each
processor must independently move from its typically low-weight
initial configuration to the high-weight region of the phase
space (``equilibration'') before measurements can start to be
taken \cite{1994:Fye,1995:Scale.Runge.Lee.Corre.Oklob.Vujic}.
Hence, while the measurement process is itself parallel,
with measurements from the
different processors combining to reduce statistical error,
equilibration is serial.
More specifically, 
let $N_{\rm ind}$ denote the number of
independent measurements required for a
desired statistical error, let $\tau_{w}$ denote the number of
Monte Carlo steps required for equilibration (``warm-up''),
let $\tau_{e}$ denote
autocorrelation time in Monte Carlo steps,
and let $N_P$ denote the number of processors in a parallel machine.
Then, for a serial or vector machine, the required number of Monte
Carlo steps $N$ is given by
\begin{equation}
  \label{eqn:serialtime}
  N = \tau_{w} + N_{\rm ind} \tau_{e}.
\end{equation}
Assuming a fluctuation-dissipation-type scenario, so that
$\tau_{w}$ and $\tau_{e}$  have similar values,
equilibration then plays a small role in the required
computational time \cite{comment:warm}.
However, for a parallel machine,
the number of steps required is 
\begin{equation}
  \label{eqn:paralleltime}
  N = \tau_{w} + N_{\rm ind} \tau_{e} / {N_{P}} .
\end{equation}
Hence, for a  massively parallel machine with thousands of processors,
equilibration provides a potential bottleneck,
which could be improved by reducing $\tau_{w}$.
It seems that such a 
bottleneck  may actually be  alleviated somewhat by  the
``sign problem'', which  can necessitate a very large value  of
$N_{\rm ind}$ for reasonable statistical error.

\subsection{Metropolis, Symmetric, and ``Generalized'' Decision  Rules}

In Monte Carlo simulations, new moves are proposed according to some procedure
and a decision is then made whether to accept
the proposed move or to reject the move and remain in the current state.
The proposal and decision together usually satisfy ``microscopic
reversibility'', or ``detailed
balance''\cite{1987:Allen.Tilde}.
The most commonly used proposal procedure stipulates that the
probability of proposing a move to state $j$ given that one is currently in
state $i$ is identical to the probability of proposing a move to
state $i$ given that one is in state $j$, though other procedures have
also been used\cite{1970:Hasti,1993:Umrig}.
Although our analysis could be generalized,
we will assume the above ``symmetric'' move
proposal procedure throughout this paper in the case
of detailed balance.

Probably  the most common rule for deciding whether or not to accept
a proposed move is the ``Metropolis decision''.
Let the Monte Carlo sampling be over the Boltzmann distribution,
let $E$ denote the energy of the current state, let $E'$ denote the energy of the proposed state,
let $\Delta E = E' - E$,
and let $\beta = 1/( k_{\rm B}T )$, where $T$ is the temperature and $k_B$ is
Boltzmann's constant.
Then, as in the original Metropolis paper\cite{1953:Metro.Rosen.Rosen.Telle},
the probability $P$ of accepting the proposed state is given by
\begin{equation}
  \label{eqn:met}
  P_{M} = \left\{
    \begin{array}{cc}
      e^{-\beta \Delta E} &, \, \Delta E > 0 \\
      1 &, \, \Delta E \leq 0 \, .
    \end{array}
  \right.
\end{equation}
Another decision rule which has also been used is the so-called ``symmetric
rule''\cite{1961:Flinn.McMan,1965:Barke},
\begin{equation}
  \label{eqn:hb}
  P_{S} = \frac{e^{-\beta \Delta E}}{1 + e^{-\beta \Delta E}} \, .
\end{equation}
We note that, since $P_{M} \ge P_{S}$ for all $\beta \Delta E$,
$P_M$ will give a higher average acceptance rate than $P_S$.

It was shown by Peskun under quite general assumptions that
the Metropolis decision would lead to smaller statistical errors in the limit
of very long simulations than would the symmetric
decision if detailed balance were satisfied \cite{1973:Pesku,1981:Pesku},
this result being most relevant to $\tau_{e}$.
The result correlates with the higher Metropolis decision
acceptance rate.
When looking at the convergence of state distributions ``operated on'' by
Monte Carlo decision rules, however, it was found that the symmetric decision
rule was superior in certain cases, particularly those where a small number of
states was accessible at each Monte Carlo step and where the quantity
$\beta \Delta E$ was typically around magnitude 1.0 or
less\cite{1976:Cunni.Meije,1977:Valle.Whitt}. This latter result
is more directly relevant to $\tau_{w}$.
Further, the Peskun analysis does not apply to certain cases of interest
which can lead to the correct limiting (e.g., Boltzmann) distribution
but which do not necessarily satisfy detailed balance,
such as systematically moving through the lattice
of a discrete system when proposing
moves as opposed to randomly selecting lattice sites\cite{1981:Pesku}.

We explore here a generalization of the Metropolis and symmetric
decisions\cite{1997:White2},
\begin{equation}
  \label{eqn:genmet}
  P = \left\{
    \begin{array}{cc}
      \frac{e^{-\beta \Delta E}}{1 + \Gamma e^{-\beta \Delta E}}
      &, \, \Delta E > 0 \\
      \frac{e^{-\beta \Delta E}}{\Gamma + e^{-\beta \Delta E}}
      &, \, \Delta E \leq 0
    \end{array}
  \right.
\end{equation}
where $\Gamma$ is a tunable parameter.  Like the standard Metropolis
and symmetric decision rules, the $P$ of Eq.~\ref{eqn:genmet}
satisfies the condition
\begin{equation}
  \label{eqn:gencon}
   e^{ - \beta E} \; P \left( E \rightarrow E^{\prime} \right) \; = \;
   e^{ - \beta E^{\prime} } \; P \left( E^{\prime} \rightarrow E \right) .
\end{equation}
When $\Gamma=0$ we recover
the Metropolis decision rule and when $\Gamma=1$ we recover the
symmetric rule.
For $0 \le \Gamma \le 1$, $P$ smoothly interpolates between the
Metropolis and symmetric limits.
However, $P$ is also defined for any $\Gamma > 1$.

\begin{figure}
  \includegraphics[angle=-90,width=\linewidth]{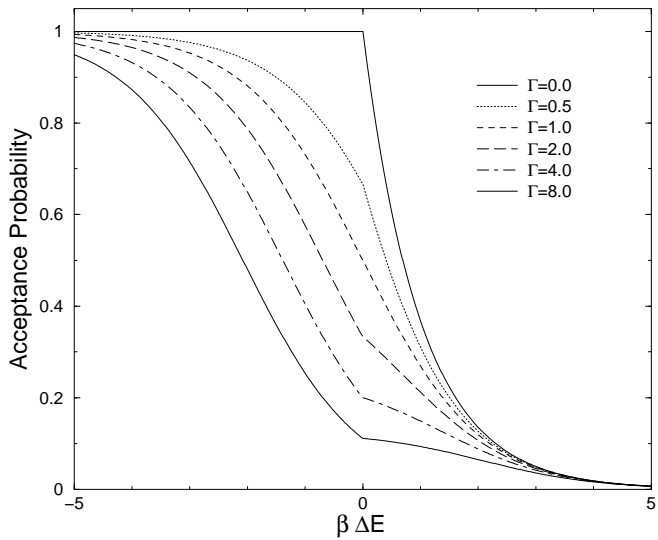}
  \caption{Effect of $\Gamma$ on acceptance probability  $P$ of Eq.~\ref{eqn:genmet}}
  \label{fig:gamma}
\end{figure}

We show in Fig.~\ref{fig:gamma} a plot of the $P$ of Eq.~\ref{eqn:genmet} versus $\beta \Delta E$ for different
values of $\Gamma$. $P$ becomes  independent of $\Gamma$
for sufficiently large magnitudes
of $\beta \Delta E$. However, it has a strong $\Gamma$ dependence around
$\beta \Delta E = 0$, with a value at $\beta \Delta E = 0$ of 1.0 when $\Gamma = 0$
(Metropolis) and 0.0 as $\Gamma \rightarrow \infty$.  Also, since $P$ is a
monotonically decreasing function of $\Gamma$ for {\it every} value of
$\beta \Delta E$, average equilibrated acceptance rates decrease monotonically
with $\Gamma$.

Two similar generalizations which interpolate between the Metropolis and symmetric
decision rules were proposed recently in the context of Monte Carlo
dynamics  by Mariz, Nobre,
and Tsallis\cite{1994:Mariz.Nobre.Tsall}.
We note that the second of these generalizations
(the so-called ``geometric unification'')
can  also  be further extended
beyond the symmetric limit, reducing $P$ when $\beta \Delta E = 0$ below the
0.5 symmetric value. 
We would expect the results which we will discuss
for the $P$ of  Eq.~\ref{eqn:genmet} to apply qualitatively
to the generalizations of Ref.~\onlinecite{1994:Mariz.Nobre.Tsall} as well.

\subsection{Auxiliary Field Quantum Monte Carlo Considerations}

The previously-cited results regarding long-run statistical
errors\cite{1973:Pesku,1981:Pesku}
can  be  extended for the $P$  of Eq.~\ref{eqn:genmet} to any  $\Gamma$, and would suggest
that the Metropolis decision rule ($\Gamma = 0$) is usually the most efficient
in that context.
However, as  mentioned in the Introduction,
there is the additional consideration
for auxiliary field quantum Monte Carlo
\cite{1981:Scala.Sugar,1981:Blank.Scala.Sugar,1981:Scala.Sugar*1,1983:Hirsc,1985:Hirsc,1986:Hirsc.Fye,1988:Fye.Hirsc} that accepting a  move is much
more computationally expensive than rejecting one, with
move acceptance dominating the computation.
Then, for greatest efficiency, one wishes to minimize not the warm-up
and equilibrated
autocorrelation times $\tau_{w}$ and $\tau_{e}$ but rather the
quantities $A_{w} \tau_{w}$ and $A_{e} \tau_{e}$,
where $A_{w}$ is an  average  move
acceptance rate during equilibration and $A_{e}$ is the
average acceptance rate after equilibration.
Even assuming that $\Gamma = 0$ (Metropolis) had the smallest $\tau$'s,
it is possible that an increase in
$\tau_{w}$ or $\tau_{e}$ could be over-compensated
by a decrease in $A_{w}$ or $A_{e}$, so that the optimal $\Gamma$
would assume some nonzero value as opposed to $\Gamma = 0$.

Specifically, we will use the Blankenbecler-Scalapino-Sugar
(BSS) quantum Monte Carlo algorithm
\cite{1981:Scala.Sugar,1981:Blank.Scala.Sugar,1981:Scala.Sugar*1}
to explore the issue of optimal $\Gamma$ with the two-dimensional
Hubbard model as a ``test Hamiltonian''. The Hubbard model interactions are
decoupled using the discrete Ising
Hubbard-Stratonovich transformation introduced
by Hirsch\cite{1983:Hirsc}. The  Monte Carlo sampling
is then performed on the Ising variables,
with effective Boltzmann weights which depend on determinants of
dense matrices resulting from
the integrating out of the fermion degrees of freedom. We note that the
usual Monte Carlo procedure (and the one which we will follow)
is to systematically move through the ``Ising lattice'' when
proposing moves, to which the Peskun analysis does not
rigorously apply\cite{1981:Pesku}.
Also, the ``step size'' for this type of simulation is fixed,
as the only possible move is a spin flip where the Ising
spin changes sign.
Hence, we do not consider in this paper effects of varying step size.

A potential additional factor
is the cost required for numerically stabilizing the BSS
algorithm \cite{1989:White.Scala.Sugar.Loh,1992:Loh.Guber,1989:Loh.Guber.Scale.Sugar},
which for a given statistical error
is proportional to $\tau_{w}$ and $\tau_{e}$ rather than
$A_{w} \tau_{w}$ and $A_{e} \tau_{e}$.
At relatively high and intermediate temperatures
this cost is negligible, but it may
constitute a significant fraction of the computational
time at very low temperatures.
In that case, one wishes rather to minimize quantities of the general
form
$\tau_{w} ( A_{w} + \kappa)$
and $\tau_{e} ( A_{e} + \kappa)$,
where the $A$'s and $\kappa$ may be of comparable size.
In the simulations which we describe, stabilization required a small
fraction of the computational time.

\section{Results}

\subsection{Procedure}

Our goal was to explore the behavior in auxiliary field quantum Monte Carlo
simulations of 
the optimal $\Gamma$ in the Monte Carlo decision rule of Eq.~\ref{eqn:genmet}.
This optimal $\Gamma$ was defined as that which led to the greatest
simulation efficiency: i.e., lowest statistical error per computational cost.
As noted previously, we estimated this efficiency
by the quantities $A_{w} \tau_{w}$ during equilibration and
$A_{e} \tau_{e}$ after equilibration, since for large systems
and typical temperatures
the leading contribution to the computational cost scales linearly
with these quantities. Again, $A_{w}$
and $A_{w}$ are average
acceptance rates during and after equilibration, respectively, and
$\tau_{w}$ is the equilibration or ``warm-up''
time and $\tau_{e}$ the autocorrelation time.

Specifically, we simulated
the two-dimensional Hubbard model\cite{1963:Hubba},
given by the Hamiltonian
\uprule
\begin{equation}
H=-t\sum_{i,j;s} \left( c^\dagger_{i,j,s} c_{i+1,j,s} + 
c^\dagger_{i,j,s} c_{i,j+1,s} + h.c.  \right) +
U \sum_{i,j} (n_{i,j,\uparrow} - {1 \over 2})
(n_{i,j,\downarrow} - {1 \over 2}) -
\mu \sum_{i,j,s} n_{i,j,s}
\label{eq:hubbard}
.
\end{equation}
\downrule
Here $c^\dagger_{i,j,s}$ creates an electron of spin $s$ at site $(i,j)$
of a two-dimensional lattice,
$n_{i,j,s} = c^\dagger_{i,j,s}c_{i,j,s}$ is
the electron occupation number for spin $s$ at site $(i,j)$,
and the hopping is between nearest neighbor sites.
We chose hopping parameter $t = 1.0$,
inverse temperature
$\beta t = t / ( k_{B} T ) = 8$,
Coulomb repulsions $U/t=4$ and $U/t=8$, and
chemical potential
$\mu / t = -1.2$ (giving an electron density per site of approximately
0.76, with $\mu = 0.0$ corresponding to half filling).
The imaginary time step was $\Delta \tau = 0.125$, so that
each Monte Carlo sweep contained
64 ``imaginary time'' slices.
All simulations were performed on a $6 \times 6$ lattice with periodic boundary
conditions.

All of the $A_{w}$ and $\tau_{w}$ and most of the $A_e$ and
$\tau_{e}$ data were obtained on the massively parallel Intel
Paragon at the Sandia National Lab MPCRL, and most of the runs were on
the full 1824 processors. Some of the $A_{e}$ and $\tau_{e}$ data
were also obtained using the CRAY T90 at the SDSC.

To obtain $A_{w}$ and $\tau_{w}$ data, each utilized node of the Paragon
was given a different random initial Ising configuration. The acceptance
rates were then collected from each node after each sweep through the
entire (space)-(imaginary time)
Ising lattice. The averaging of the data from all the nodes increased the
signal-to-noise ratio to the extent that fits could then  be made to the
decay of the acceptance rate from its initial  higher value $A_0$,
when moves are more likely
due to the high probability of being in a low-weight state,
to the lower equilibrated value $A_e$. $A_w$ was
calculated as the average of the
initial acceptance rate $A_0$ and the equilibrated rate $A_e$.
Several observables $\cal O$
($n = n_\uparrow + n_\downarrow$,
$\sigma_{z} = n_\uparrow - n_\downarrow$,
the staggered spin structure functions
$S_{zz}(\pi,\pi)$ and $S_{xx}(\pi,\pi)$, and kinetic and total
energies \cite{1985:Hirsc,1986:Hirsc})
were monitored similarly to the acceptance rate,
and  it was found that they all equilibrated at least as
quickly as did the acceptance rate itself.
Fits of the form
$A_{e} +  (A_{0} - A_{e}) \, exp(-\tau / \tau_{w})$
were made to the acceptance rate
data, defining the $\tau_{w}$'s.
Typically, only data up to twenty sweeps was used in this fit,
as the signal became lost in the noise for a larger number
of sweeps.

A somewhat analogous procedure was followed in
computing the $\tau_{e}$'s.
First, for the observables $\cal O$ listed above, autocorrelation
functions $C(\tau)$ were calculated from the
equilibrated data using the
formula
\begin{equation}
  \label{eqn:autocorr}
    C(\tau) = \frac{\langle {\cal O}_i {\cal O}_{i+\tau} \rangle -
    \langle {\cal O}_i \rangle^2}{\langle {\cal O}_{i}^{2} \rangle -
    \langle {\cal O}_i \rangle^2},
\end{equation}
where $\tau$ is the lag time.
The autocorrelations were then fit to the
form $exp( - \tau  / \tau_{e})$,
which we here took to define
the equilibrated  $\tau_{e}$ autocorrelation
times \cite{1994:Kawas.Guber.Evert2}.
Better statistics could be obtained for the  $\tau_{e}$'s
due to the existence of more equilibrated than
equilibrating (warm-up) data.

Of those considered, it was found that 
the observable with the longest calculated
autocorrelation time, $\tau_{e}$, was $ n_\uparrow - n_\downarrow $.
Hence, we will focus primarily on the $\tau_{e}$ behavior of
$ n_\uparrow - n_\downarrow $.

\subsection{Warm-Up Results}
 
\begin{figure}
    \includegraphics[angle=-90,width=\linewidth]{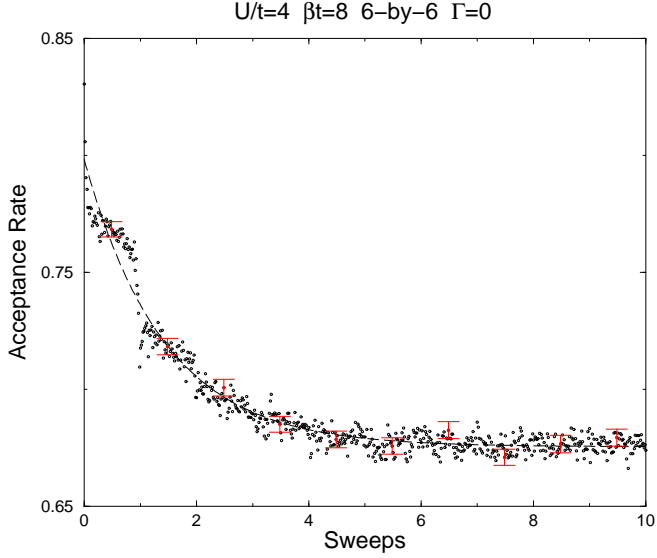}
    \caption{The acceptance rate
    for $\Gamma = 0$ and $U/t=4$, shown starting from random configurations
    at step 0 (one sweep contains 64 imaginary time steps)
    until it equilibrates.
    For clarity, error bars are shown only once every 64 points,
    but all error bars are of the same general magnitude.
    The points are averaged data from 1824 processors.
    The dashed line
    shows a fit of the function $0.675+0.123 \exp(- \tau /1.42)$}
    \label{fig:accept_warm}
\end{figure}

In Fig.~\ref{fig:accept_warm} we show the equilibration of the acceptance
rate during warm-up for $\Gamma = 0$ (Metropolis decision)
and $U/t=4$ using data
averaged from 1824 processors. We also show an exponential decay
fit to the relaxation of the acceptance rate from its initial
to its equilibrated value.

\begin{figure}
  \includegraphics[angle=-90,width=\linewidth]{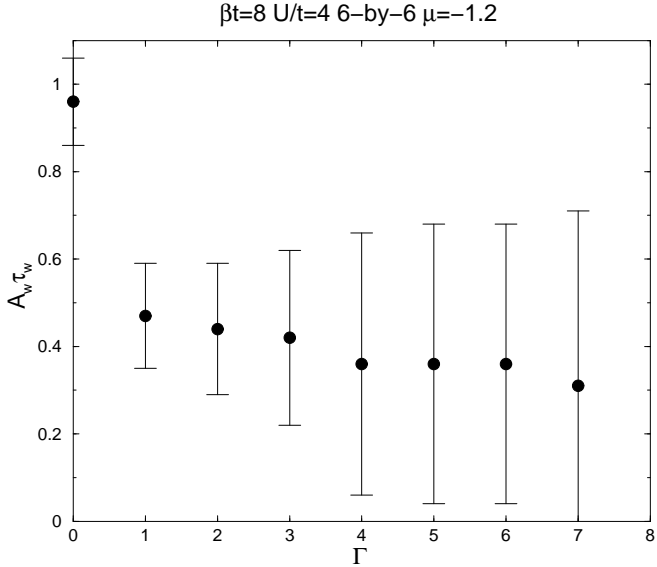}
  \caption{$A_{w} \tau_w$ versus $\Gamma$ for $U/t = 4$.}
  \label{fig:atauw}
\end{figure}

Such fits for various values of $\Gamma$ were used
to obtain the $A_{w} \tau_{w}$ results for $U/t=4$ of Fig.~\ref{fig:atauw},
where again $A_{w} = 1/2 \, ( A_{0} + A_{e} )$
is an average warm-up acceptance rate.
(The error bars for $U/t=8$ were too large to observe
statistically significant differences.)
Note the drop shown in Fig.~\ref{fig:atauw}
in $A_{w} \tau_{w}$
when going from $\Gamma = 0$ (Metropolis decision rule) to
$\Gamma = 1$ (``symmetric'' rule), suggesting increased
efficiency.

\subsection{Equilibrated Results}

\begin{figure}
    \includegraphics[angle=-90,width=\linewidth]{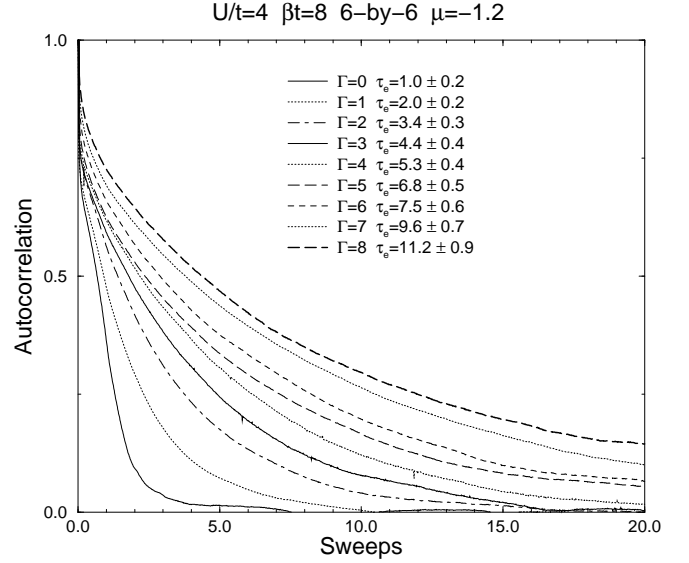}
    \caption{Autocorrelation of $ n_\uparrow -
      n_\downarrow $ for various values of $\Gamma$ for $U/t =
      4$. These
      curves were fit to the function $\exp(-\tau /\tau_e)$ to
      obtain $\tau_e$.}   
    \label{fig:autocorr}
\end{figure}

\begin{figure}
    \includegraphics[angle=-90,width=\linewidth]{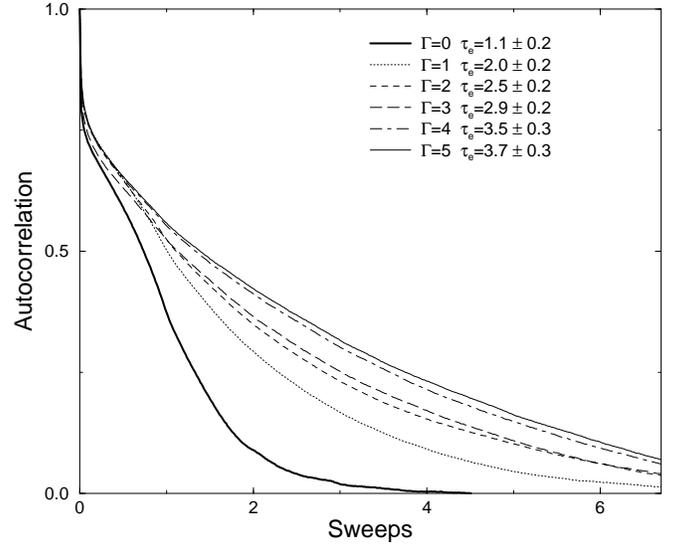}
    \caption{Autocorrelation of $ n_\uparrow -
      n_\downarrow $ for various values of $\Gamma$ for $U/t = 8$. These
      curves were fit to the function $\exp(-\tau /\tau_e)$ to
      obtain $\tau_e$.}   
    \label{fig:autocorru8}
\end{figure}

In Fig.~\ref{fig:autocorr} and Fig.~\ref{fig:autocorru8},
we show data for the autocorrelation $C( \tau )$ of
Eq.~\ref{eqn:autocorr} with operator
${\cal O} = n_\uparrow - n_\downarrow$.
The data were obtained by taking samples of $n_\uparrow -
n_\downarrow$ 
at every
time slice for a large number of Monte Carlo sweeps through the lattice.
Also given are Monte Carlo autocorrelation
times, $\tau_e$, obtained by an exponential fit, as well as the
corresponding error bars.
The $\tau_e$'s and errors were calculated with  the use of a Fourier
transform method \cite{spectrum}.
As might be expected from the lower acceptance rates for larger
$\Gamma$, the $\tau_e$ autocorrelation times increase for
larger $\Gamma$.
 
\begin{figure}
    \includegraphics[angle=-90,width=\linewidth]{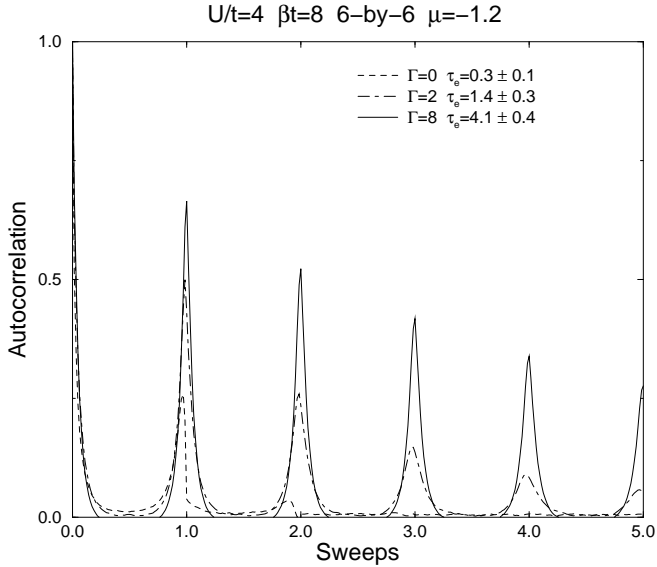}
    \caption{Autocorrelation of $S_{zz} (\pi,\pi)$
      for various values of $\Gamma$ for $U/t=4$.  Taking just the peaks,
      these curves can also be fit to $\exp(- \tau /\tau_e)$
      to obtain a $\tau_e$ for this observable.}
    \label{fig:autocorr.afzz}
\end{figure}

As shown in Fig~\ref{fig:autocorr.afzz}, we observed a
``ringing'' in the autocorrelation for the staggered
spin structure function in the $z$-direction, $S_{zz} (\pi,\pi)$.
No such ringing was observed for any other observable studied,
including the staggered
spin structure function in the $x$-direction, $S_{xx} (\pi,\pi)$.
(It is known that a BSS simulation with
discrete Hubbard-Stratonovich decoupling can lead to
different variances for quantities with the same averages
which are measured in different directions \cite{1986:Hirsc}.)

\begin{figure}
  \includegraphics[angle=-90,width=\linewidth]{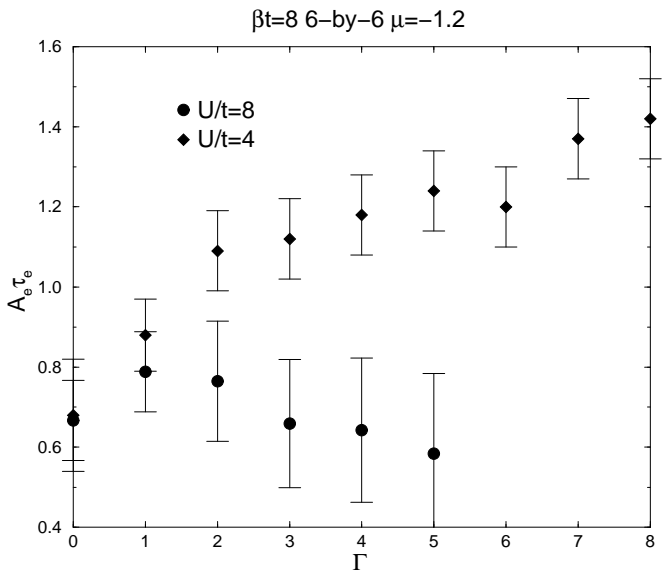}
  \caption{$A_{e} \tau_{e}$ versus $\Gamma$ for $U/t=4$ and $U/t=8$.}
  \label{fig:atauc}
\end{figure}

In Fig.~\ref{fig:atauc}, we show $A_{e} \tau_{e}$
for both $U/t = 4$ (top) and $U/t = 8$ (bottom).
There were no statistically significant changes observed
in  $A_{e} \tau_{e}$ for $U/t = 8$ in the studied range
$0 \le \Gamma \le  5$.
However, for $U/t = 4$, $A_{e} \tau_{e}$ was observed to
increase with increasing $\Gamma$, and a monotonic increase
is quite consistent with the data. Of the $0 \le \Gamma \le 8$
values studied
for $U/t = 4$, $\Gamma = 0$ gave the (statistically significant)
lowest value of $A_{e} \tau_{e}$, suggesting that the
Metropolis decision was most efficient in this case.

\section{Discussion}

We have explored possible improvements in the efficiency of
determinantal fermion Monte Carlo simulations
using a generalization of the standard Metropolis
decision rule,
\begin{equation}
  \label{eqn:genmetp}
  P = \left\{
    \begin{array}{cc}
      \frac{e^{-\beta \Delta E}}{1 + \Gamma e^{-\beta \Delta E}}
      &, \, \Delta E > 0 \\
      \frac{e^{-\beta \Delta E}}{\Gamma + e^{-\beta \Delta E}}
      &, \, \Delta E \leq 0 \, .
    \end{array}
    \right.
\end{equation}
This generalization interpolates between the
Metropolis \cite{1953:Metro.Rosen.Rosen.Telle}
and so-called ``symmetric''\cite{1961:Flinn.McMan,1965:Barke}
decision rules but also has additional flexibility:
the average acceptance rate can be smoothly
reduced from the (maximal) Metropolis value (when $\Gamma = 0$)
through the symmetric
value ($\Gamma = 1$) down to an arbitrarily small value.
Specifically, we performed auxiliary field quantum Monte
Carlo simulations \cite{1981:Scala.Sugar,1981:Blank.Scala.Sugar,1981:Scala.Sugar*1,1983:Hirsc,1985:Hirsc,1986:Hirsc.Fye,1988:Fye.Hirsc,1989:White.Scala.Sugar.Loh,1992:Loh.Guber};
however, our approach and results may have application
to electronic structure fermion Monte Carlo methods as
well \cite{1994:Hammo.Leste.Reyno,1995:Ander,1996:Ceper.Mitas,1996:Mitas,1998:Night.Umrig,1999:Umrig2}.

One might in general expect that a lower acceptance rate would be less
computationally efficient, since the sampling would then proceed
more slowly through the configuration space.
However, determinantal quantum Monte Carlo simulations
have the feature that an accepted move is generally much more
costly than a rejected one and that the calculations associated
with move acceptance will typically dominate the computation.
Then, instead of trying to
minimize autocorrelation times $\tau_e$ for greater efficiency,
one rather wishes to minimize $A_{e} \tau_{e}$, where $A_e$ is the
average equilibrated acceptance rate.
A similar argument holds for $A_{w} \tau_w$ as opposed to $\tau_{w}$,
where $\tau_w$ is the equilibration or ``warm-up'' time and
$A_{w}$ is the averaged acceptance rate during the equilibration
process. Since equilibration poses a potential (serial)
bottleneck in parallelization, we considered the behavior
both of $A_{e} \tau_{e}$ and of $A_{w} \tau_{w}$.
In particular, we explored whether generalizations of the
Metropolis decision
could reduce $A_{e} \tau_{e}$ or $A_{w} \tau_{w}$.

For a test system, we utilized the two-dimensional Hubbard
model at approximately 3/4 filling
with $U/t=4$ (weak coupling) and $U/t=8$
(moderate to strong coupling).
For computing an autocorrelation time $\tau_e$, we used
the observable $n_{\uparrow} - n_{\downarrow}$, which had the
longest such calculated
autocorrelation time of any of the observables considered.
For calculating the ``warm-up'' time $\tau_w$, we
monitored how the acceptance rate equilibrated
from random initial configurations.
When $U/t=4$, both $\tau_e$ and $A_{e} \tau_e$
increased with $\Gamma$, suggesting that $\Gamma = 0$
(Metropolis decision) was the most efficient.
The $\tau_e$'s grew monotonically with $\Gamma$ when
$U/t=8$, but there was no statistically significant variation
observed in $A_{e} \tau_{e}$ for the $\Gamma$'s tested.
The error bars in $A_{w} \tau_{w}$ when $U/t=8$ were too
large for meaningful comparisons;
however, when $U/t=4$, $A_{w} \tau_{w}$ was approximately halved in
going from $\Gamma = 0$ (Metropolis) to $\Gamma = 1$
(symmetric), indicating that the symmetric decision was more
efficient during equilibration.

A byproduct of this work was
calculations of autocorrelation times
$\tau_e$ and ``warm-up'' times $\tau_w$ for
some sample auxiliary field (two-dimensional Hubbard model)
quantum Monte Carlo runs, which calculations
to the best of our knowledge have not previously appeared.
For both Metropolis and symmetric decision rules,
the $\tau$'s were typically at most a few sweeps through
the Ising (space)-(imaginary-time) auxiliary field lattice.
These values are shorter than might have been expected,
and suggest that the often several thousand warm-up sweeps
conventionally done in vector simulations are very adequate.
Such $\tau$ values may also provide useful ``ball park'' estimates
when planning parallel runs.

It was clear from our simulations that which of the
Monte Carlo decisions
is most efficient in determinantal fermion Monte Carlo
can depend on specifics of the model and parameters.
However, the standard Metropolis decision 
was optimal or near optimal
once equilibration had been reached
in the two sample cases studied,
suggesting that it is efficient after warm-up.
The symmetric decision, however, was more efficient
{\em during} warm-up.  Particularly if
the (serial) warm-up process becomes a bottleneck in
parallel simulations, this suggests that the use of
symmetric or other non-Metropolis decision rules could lead to
greater computational speed.

\section*{Acknowledgements}

The authors are grateful to D.J. Scalapino and
R.L. Sugar for helpful discussions and comments.
The work of  CLM was funded by the US Department
of Energy under Grant No. DE--FG03--85ER45197.
The work of RMF was supported by the US DOE MICS program under
Contract DE-ACO4-94AL8500. Sandia is a multiprogram laboratory operated
by Sandia Corporation, a Lockheed Martin Company, for the US DOE.
This work was made possible by computer time on
the Paragon at the Sandia National Labs MPCRL and the Cray T90 at NPACI.

\bibliography{chris,monte}

\begin{thebibliography}{10}
\bibitem{comment:warm}
In some types of simulations, equilibration can be more costly simply due to
  the fact that more Monte Carlo steps are required to traverse the distance
  from a typical initial configuration to the high-weight region of the phase
  space than to move around the high-weight region itself. Also, the time
  required to equilibrate can sometimes be reduced by starting with
  higher-weight initial configurations, if they can be determined.
\end{thebibliography}
\bibliographystyle{prsty_chris}

\end{multicols}

\end{document}